# High-Resolution Transmission Spectra of Earth through Geological Time

Lisa Kaltenegger[1,2], Zifan Lin[1,2] & Jack Madden[1,2]
[1]Cornell University, Astronomy and Space Sciences Building, Ithaca, NY 14850, USA
[2]Carl Sagan Institute, Space Science Building 311, Ithaca, NY 14850, USA

ABSTRACT
The next generation of ground- and space-based telescopes will be able to observe rocky Earth-like planets in the near future, transiting their host star. We explore how the transmission spectrum of Earth changed through its geological history. These transmission spectra provide a template for how to characterize an Earth-like exoplanet – from a young prebiotic world to a modern Earth. They also allow us to explore at what point in its evolution a distant observer could identify life on our *Pale Blue Dots* and other worlds like it. We chose atmosphere models representative of five geological epochs of Earth's history, corresponding to a prebiotic high $CO_2$-world 3.9 billion years ago (Ga), an anoxic world around 3.5 Ga, and 3 epochs through the rise of oxygen from 0.2% to present atmospheric levels of 21%.

Our transmission spectra show atmospheric spectral features, which would show a remote observer that Earth had a biosphere since about 2 billion years ago. These high-resolution transmission spectral database of Earth through geological time from the VIS to the IR is available online and can be used as a tool to optimize our observation strategy, train retrieval methods, and interpret upcoming observations with JWST, the Extremely Large Telescopes and future mission concepts like Origins, HabEx, and LUOVIR.

## INTRODUCTION

Among the more than 4000 discovered exoplanets to date are dozens of Earth-size planets (see, e.g., Udry et al. 2007; Borucki et al. 2011, 2013; Kaltenegger & Sasselov 2011; Batalha et al. 2013; Kaltenegger et al. 2013; Quintana et al. 2014; Torres et al. 2015), including several with similar irradiation to Earth (see e.g., Kane et al 2016, Kaltenegger 2017, John et al 2018, Berger et al 2018).

The space-based James Webb Space telescope (JWST) is scheduled to launch in early 2021 and several ground-based Extremely Large Telescopes (ELTs) are currently under construction or in planning, like the Giant Magellan Telescope (GMT), the Thirty Meter Telescope (TMT), and the Extremely Large Telescope (ELT), which are designed to be able to undertake the first measurements of the atmospheres of Earth-sized planets (see e.g. Kaltenegger & Traub 2009, Kaltenegger et al 2010, Garcia-Munoz et al 2012, Hedelt et al. 2013, Snellen et al 2013, Rodler & Lopez-Morales 2014, Betremieux & Kaltenegger 2014, Misra et al 2014, Stevenson et al 2016, Barstow et al 2016, Kaltenegger et al 2019, Zin & Kaltenegger 2019). Several future mission concepts like Origins (Battersby et al 2018), Habex (Mennesson et al. 2016) and LUOVIR (LuvoirTeam et al 2018) are currently being designed to be able to explore the atmospheric composition of Earth-sized planets.

Earth's atmosphere has undergone a substantial evolution since formation (see e.g. Walker 1977, Zahnle et al 2007, Lyones et al 2014). Previous work by one of the authors modeled Earth's reflection and emission spectra through geological time, representative of exoplanet observations seen as directly imaged *Pale Blue Dots* (Kaltenegger et al. 2007). A second paper including one of the authors investigated how the reflection and emission spectra of Earth through its geological history from anoxic to modern Earth-like planets changes if they are orbiting different Sun-like host stars from F0V to M8V spectral type (Rugheimer & Kaltenegger 2018). The surface UV environment for Earth through geological time for our Sun and around different Sun-like host stars shows comparable UV surface environments for such planets as discussed in Rugheimer & Kaltenegger (2018) and O'Malley-James & Kaltenegger (2017, 2019).

While emission and reflection spectra for models of Earth through geological time exist (e.g. Kaltenegger et al 2007, Rugheimer & Kaltenegger 2018), transmission spectra have been focused on modern Earth so far (see e.g., Ehrenreich et al. 2006, Kaltenegger & Traub 2009, Palle et al. 2009, Vidal-Madjar et al. 2010, Rauer et al. 2011, García Munoz et al. 2012, Hedelt et al. 2013, Betremieux & Kaltenegger 2013, 2014, Misra et al. 2014).

Here we model the high-resolution transmission spectra for five geological epochs in Earth's history (Table 1): representative of a high-$CO_2$ prebiotic world as epoch 1 around 3.9 Ga, an Anoxic world as epoch 2 around 3.5 Ga and 3 epochs during the rise of oxygen, corresponding to the timeframe of the rise of oxygen in Earth's atmosphere between about 2.4 Ga to today (see review by Lyons et al 2014). We modeled epoch 3 after the Grand Oxygenation Event, with 1% PAL (present atmospheric levels) of $O_2$, epoch 4 after the Neoproterozoic Oxygenation Event, with 10% PAL $O_2$. Epoch 5 represents modern Earth atmosphere with 21% $O_2$. We use a solar evolution model (Claire et al. 2012) to establish the incident Solar Flux through Earth's geological evolution.

Our high-resolution database of transmission spectra from the visible to the Infrared (0.4 μm to 20 μm) for Earth through geological time, which is freely available online (www.carlsaganinstitute.org/data), is a tool to enable effective observations and first interpretation of atmospheric spectra of Earth-like planets, using our planet's evolution as a template. Our models and transit spectra include climate indicators like $H_2O$ and $CO_2$ as well as biosignatures like the combination of $O_2$ or $O_3$ in combination with a reduced gas like $CH_4$ (see discussion on biosignatures e.g. in Kaltenegger 2017). The term biosignatures is used here to mean remotely detectable atmospheric gases that are produced by life and are not readily mimicked by abiotic processes, e.g. the $CH_4$ + $O_2$ (Lederberg 1965, Lovelock 1965) or $CH_4$ + $N_2O$ (Lippincott et al. 1967) pairs.

In section 2 we describe our model. In section 3 we present the transmission spectra for 5 epochs in Earth's geological history and discuss the absorption features of climate indicators and biosignatures in low and high-resolution from the visible to infrared wavelength. Section 4 discusses and summarizes our results.

**METHODS**

We used EXO-Prime (for details see Kaltenegger et al 2007, Kaltenegger & Sasselov 2010 and Madden & Kaltenegger 2019) to simulate Earth's atmosphere and transmission spectra for 5 geological epochs. EXO-Prime is a coupled 1D iterative climate-photochemistry code (see e.g. Kasting & Ackerman 1986, Pavlov &

Kasting 2002, Segura et al. 2005, 2007, Haqq-Misra et al. 2008, Arney et al. 2016, Madden & Kaltenegger 2019), with a line by line Radiative Transfer code (e.g. Traub & Stier 1976; Kaltenegger & Traub 2009) for rocky exoplanets, which was originally developed for Earth. EXO-Prime has been validated for visible to infrared wavelengths by comparison to Earth observed as an exoplanet from different missions like EPOXI, the Mars Global Surveyor, Shuttle data, and multiple earthshine observations (Kaltenegger et al. 2007; Kaltenegger & Traub 2009; Rugheimer et al. 2013).

We calculate the high-resolution transmission spectra at a resolution of 0.01 cm$^{-1}$ using opacities from the 2016 HITRAN database (Gordon et al. 2017) for $O_2$, $O_3$, $H_2O$, $CO_2$, $CH_4$, $N_2O$, $CH_3Cl$, $SO_2$, $H_2S$, $H_2O_2$, $OH$, $HO_2$, $HOCl$, $H_2CO$, $HCl$, $ClO$, $NO_2$, $NO$, $HNO_3$, and $CO$. For $CFCl_3$ (Sharpe et al. 2004) and $N_2O_5$ (Wagner & Birk 2003) we use cross-sections. We include $CO_2$ line mixing (see also Niro et al. 2005a, 2005b). For $CO_2$, $H_2O$, and $N_2$, we use measured continua data instead of line-by-line calculations in the far wings (see Traub & Jucks 2002).

Earth's atmosphere cannot be probed in primary transit below 12 km, because refraction from the deeper atmospheric regions deflects light away from a distant observer in an Earth–Sun geometry (see e.g. Garcıa Munoz et al. 2012, Betremieux & Kaltenegger 2014, Misra et al. 2014). Our transmission spectra show the cutoff due to refraction at 12km for all epochs. Clouds do not significantly affect the strengths of the spectral features in Earth's transmission because most clouds on Earth are located at altitudes below 12km.

We base the transmission spectra for Earth through geological times on atmosphere models by Kaltenegger et al. (2007) and Rugheimer et al. (2018): The models for each epoch are discussed in detail in those two papers and summarized in Table 1 and Figure 1. The changing solar constant accounts for the lower solar incident flux at earlier times in Earth's history, following the prediction of a 30% reduction in solar flux for a young Earth at 4.6 Ga (Claire et al 2012). All epochs assume a 1 bar surface pressure, consistent with geological evidence for paleo-pressures close to modern values (e.g. Somet al. 2012; Marty et al. 2013).

Table 1: Chemical mixing ratios for major atmospheric gases in our model atmospheres, for 5 epochs through Earth's geological history from prebiotic to anoxic atmospheres representative of 3.9 Ga and 3.5 Ga in Earth's history to 3 models which capture the rise of oxygen from 0.01PAL $O_2$ to modern Earth with 21% $O_2$.

| Time Period | Solar Constant | Epoch | $CO_2$ | $CH_4$ | $O_2$ | $O_3$ | $N_2O$ |
|---|---|---|---|---|---|---|---|
| now | 1.00 | 5 | 3.65E-04 | 1.65E-06 | 2.10E-01 | 3.00E-08 | 3.00E-07 |
| 0.5 - 0.8 | 0.95 | 4 | 1.00E-02 | 4.15E-04 | 2.10E-02 | 2.02E-08 | 9.15E-08 |
| 1.0 - 2.0 | 0.87 | 3 | 1.00E-02 | 1.65E-03 | 2.10E-03 | 7.38E-09 | 8.37E-09 |
| 3.5 | 0.77 | 2 | 1.00E-02 | 1.65E-03 | 1.00E-13 | 2.55E-19 | 0 |
| 3.9 | 0.75 | 1 | 1.00E-01 | 1.65E-06 | 1.00E-13 | 2.55E-19 | 0 |

Epoch 1 is a $CO_2$-rich atmosphere of a prebiotic world, representative of Early Earth around 3.9 Ga. Epoch 2 is an Archaen world, representative of a young Earth around 3.5 Ga. Epoch 3 corresponds to a Paleo- and Meso-proterozoic Earth (about 2 to 1 Ga), when oxygen started to rise in Earth's atmosphere. We use 0.21% $O_2$ (1% PAL) for this model. Epoch 4 corresponds to the proliferation of multicellular life on Neoproterozoic Earth (about 0.8 to 0.5 Ga) when the oxygen concentration had risen to 10% PAL (2.1% $O_2$). Epoch 5 corresponds to modern Earth with 21% $O_2$. Note that the time of the oxygen rise has recently been moved to a later stage in Earth's evolution (see e.g. review in Lyones et al 2014), which is reflected in the time ranges gives in Table 1 for epoch 3 and epoch 4, instead of geological times given in our earlier paper (Kaltenegger et al. 2007), which based $O_2$ concentrations on work by Holland et al. (2006).

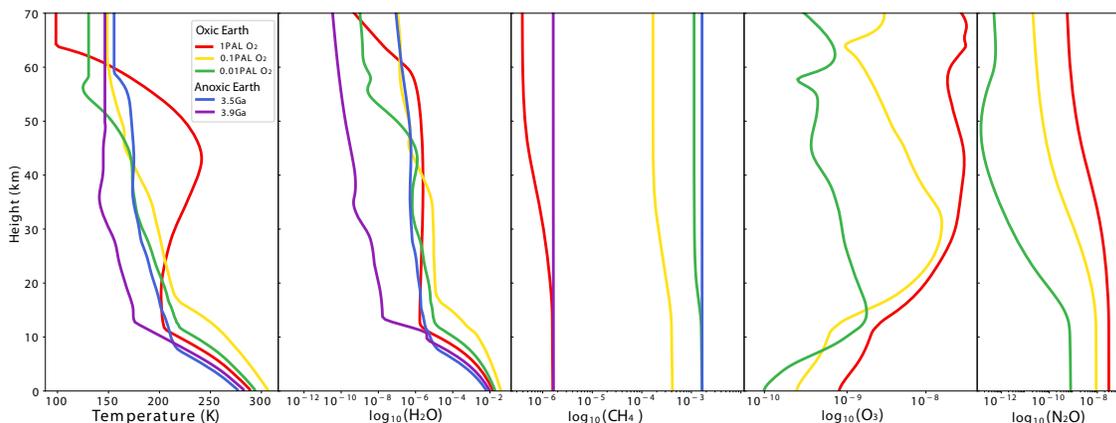

Fig.1: Temperature and mixing ratios for major atmospheric gases in our model atmospheres, representative of 5 epochs through Earth's geological evolution from an $CO_2$-rich prebiotic atmosphere for Earth around 3.9Ga to an anoxic atmosphere around 3.5Ga and 3 models, which capture the rise of oxygen from 0.01PAL $O_2$ to 1PAL (21% $O_2$) on modern Earth. The mixing ratios shown from left to right are for $H_2O$, $CH_4$, $O_3$ and $N_2O$ (see also Table 1).

The transmission spectra in figures 2 are smoothed with a triangular kernel to a resolving power of 700 for clarity. We did not add noise to the spectra to provide theoretical input spectra for several instruments, which all have different instrument-specific noise profiles which can easily be added to our model to provide realistic observation simulations.

## RESULTS

All transmission spectra are available online for a minimum resolution of $\lambda/\Delta\lambda > 100,000$ for the full wavelength range from 0.4µm to 20µm (0.01cm$^{-1}$ steps). Figure 2 shows the transmission spectra for a resolution of $\lambda/\Delta\lambda = 700$ for clarity, with the most prominent spectral features identified. The five atmosphere models are representative of Earth through geological time sorted from modern Earth on top to Early Earth on the bottom:

Modern Earth's transmission spectrum is shown as the top row (Epoch 5, 21% $O_2$), followed by the transmission spectrum for Neoproterozoic Earth (Epoch 4, 0.5 to 0.8Ga, $O_2$ = 2.1 x 10$^{-2}$ (10% PAL)), followed by the transmission spectrum for a Paleo- and Meso-proterozoic Earth (Epoch 3, 1 to 2Ga, $O_2$ = 2.1 x 10$^{-3}$ (1% PAL)), an anoxic Earth (Epoch 2, 3.5 Ga) and a prebiotic Earth (epoch 1, 3.9 Ga), which is shown in the bottom row.

Throughout the atmospheric evolution of our Earth model, different absorption features dominate Earth's spectrum with $CH_4$ and $CO_2$ being dominant in Early Earth models, where they are more abundant and $O_2$ and $O_3$ features increasing with $O_2$ abundance from a Paleo- and Meso-proterozoic Earth in epoch 3 to modern Earth in epoch 5 (Table 1).

Note that several features overlap at the resolution of $\lambda/\Delta\lambda = 700$, which is shown in Fig. 2 for clarity and are not specifically labeled. In the high-resolution online transmission spectra individual spectral lines can be easily discerned for these molecules, as shown in Fig. 3 for $O_2$ and $O_3$ for biotic atmospheres (epoch 5 to epoch 3), for a resolution of $\lambda/\Delta\lambda = 100,000$ as proposed for several instruments on the ELTs.

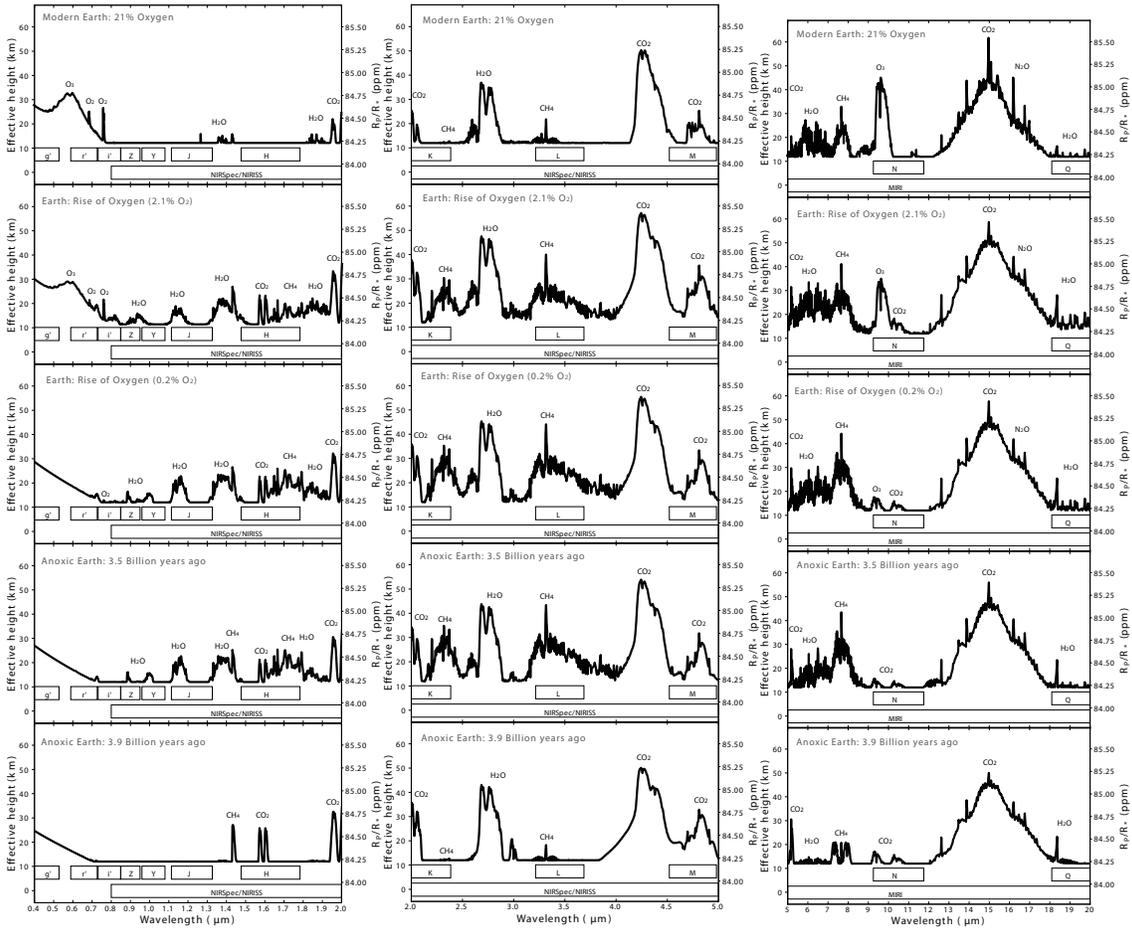

Fig. 2: Model Spectra for Earth through geological time from 0.4 to 20 μm shown at a resolution of $\lambda/\Delta\lambda = 700$ for 5 epochs through Earth's geological time from an anoxic atmosphere 3.9 Ga to an anoxic atmosphere with less $CO_2$ and $CH_4$ around 3.5 Ga and 3 models which capture the rise of oxygen from 0.01 PAL $O_2$ to 1PAL (21% $O_2$) on modern Earth, which started around 2.4 billion years ago (Ga).

**Absorption features in the visible wavelength range** (left panel, 0.4 to 2 μm): Modern Earth (epoch 5, top row) shows a strong absorption feature for $O_2$ at 0.76 μm, with a weaker feature at 0.69 μm. $O_3$ shows a broad feature, from approximately 0.45 to 0.74 μm. These features decrease with decreasing oxygen composition for younger Earth models and disappear for anoxic Earth atmosphere models (epoch 1 and epoch 2). $CH_4$ in modern Earth's atmosphere shows no significant visible absorption features in transmission in the visible in Fig. 2, but at higher abundance, it shows absorption features at 1.7, and also at 0.88 and 1.04 μm in early Earth models. H$_2$O shows absorption features at a wide range of wavelength in the visible at 0.73, 0.82, 0.95, and 1.14 μm. CO$_2$ does not show visible features at present abundance, but in a high-CO$_2$ atmosphere of 10% CO$_2$, as in early Earth evolution stages, the weak 1.06, 1.3, 1.6 and 2μm features can be seen in Figure 2.

**Absorption features in the NIR wavelength range** (middle panel from 2 to 5 μm): Neither O$_2$ nor O$_3$ show absorption features in the NIR. CH$_4$ shows absorption features at 2.4 and 3.3 μm, which increase with CH$_4$ abundance in Earth's atmosphere for earlier geological epochs. Several CO$_2$ features can be identified in Fig. 2 with increasing CO$_2$ abundance for younger Earth models. H$_2$O abundance and absorption feature strength increase with increasing surface temperature and consequence evaporation rate for epoch 3 and epoch 4.

**Absorption in the IR wavelength range** (right panel from 5 to 20 mm): At 9.6 μm the strength of the absorption feature of O$_3$ decreases with decreasing O$_3$ abundance for younger oxic Earths. It is a saturated feature and therefore an excellent qualitative but poor quantitative indicator for the existence of O$_2$. A smaller O$_3$ feature can be seen at 9μm also decreasing in strength with decreasing O$_3$ abundance. For anoxic atmospheres the O$_3$ feature is not visible. At 7.6 μm the absorption feature of CH$_4$ becomes stronger with increasing CH$_4$ abundance for younger Earth models. The main CO$_2$ feature at 15μm as well as several smaller CO$_2$ features at 10.4 and 9.4 μm increase with increasing CO$_2$ abundance for earlier Earth models. H$_2$O features can be seen at 7 and 20 μm. N$_2$O has spectral features at 16.89 μm, which can be seen in epoch 3 and 4 in the wing of the 15 μm CO$_2$ feature, with smaller features at 7.75, 8.52, and 10.65 μm, which overlap with other spectral features at the resolution shown in Fig. 2.

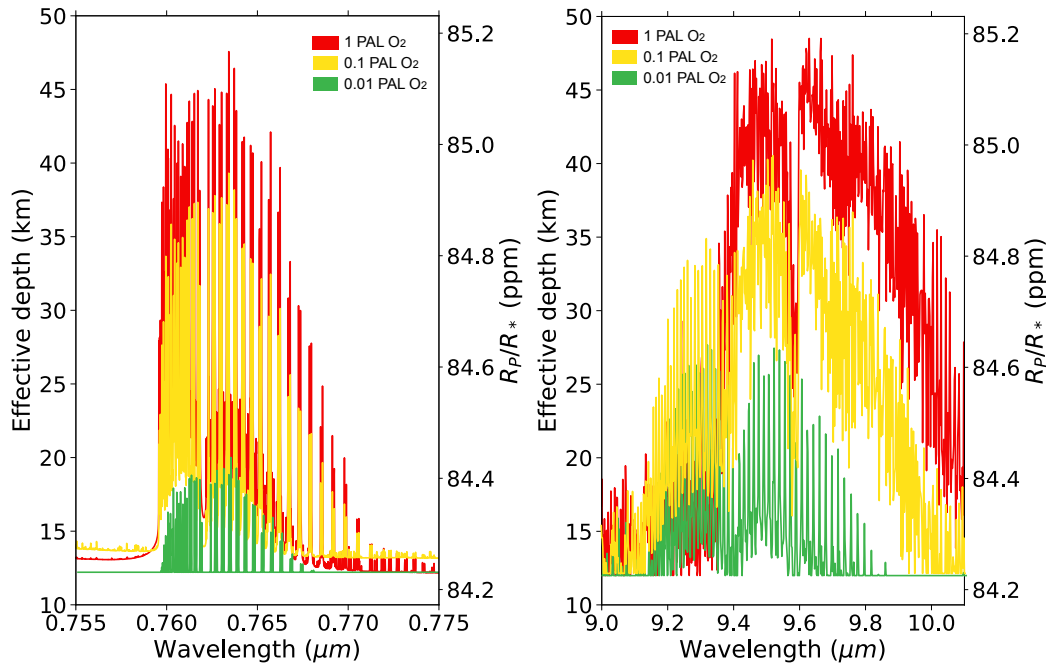

Fig. 3: High-resolution ($\lambda/\Delta\lambda$ > 100,000) for the 0.76 μm O$_2$ and 9.6 μm O$_3$ feature for the rise of oxygen from 0.01 to the present atmospheric level (PAL) of oxygen, which is 21% in Earth's modern atmosphere.

Note that in the online high-resolution transmission spectra some absorption features, which are not apparent in Figure 2 at a resolution of $\lambda/\Delta\lambda = 700$, can be identified. As an example we show the change in both the $O_2$ feature at 0.76 μm and the $O_3$ feature at 9.6 μm through geological time in Figure 3 for a minimum resolution of $\lambda/\Delta\lambda = 100,000$ for the whole wavelength range shown.

**Discussion and Conclusion**

We generated a high-resolution transmission spectral database of atmospheric models representative of Earth through its geological history from the VIS to the IR (0.4 to 20μm) with a minimum resolution of 100,000. These transmission spectra provide a template of how to remotely characterize Earth-like exoplanets with upcoming ground- and space-based telescopes and explore at what point in its evolution a distant observer could identify life on our own planet and others like it.

We chose atmospheres representative of five geological epochs of Earth's history, corresponding to a prebiotic high $CO_2$-world and an anoxic world at 3.9 and 3.5 billion years ago, as well as 3 epochs through the rise of $O_2$ from 1% of to present atmospheric levels, which started 2.4 billion years ago on Earth. Throughout the atmospheric evolution of our Earth, different absorption features dominate Earth's transmission spectrum (shown in Fig. 2 at a resolution of $\lambda/\Delta\lambda = 700$) with $CH_4$ and $CO_2$ being dominant in Early Earth models, where they are more abundant. $O_2$ and $O_3$ spectral features become stronger with increasing abundance during the rise of oxygen (Epoch 3 to 5).

Analyzing the emergent spectrum of Earth, taken by the Galileo probe, Sagan et al. (1993) concluded that the large amount of $O_2$ in the presence of $CH_4$ is strongly suggestive of biology, as Lovelock (1965) and Lederberg (1965) had suggested earlier. On short timescales, the two species react to produce $CO_2$ and $H_2O$ and therefore, must be constantly replenished to maintain detectable concentrations (see e.g. review Kaltenegger 2017). It is their quantities and detection along with other atmospheric species in the planetary context that solidify a biological origin (as discussed in detail in several recent reviews e.g. Kasting et al. 2014, Kaltenegger 2017, Swieterman 2018). On Earth, the combination of these gases can be detected in several different wavelengths, dependent on their abundance and Earth's geological evolution, as shown in Fig. 2. The strongest features in the visible are $O_2$ (0.76 μm) and $O_3$ (0.6 μm), in the NIR $CH_4$ (2.4μm), and in the thermal IR $O_3$ (9.6μm) and $CH_4$ (7.6 μm). High-resolution spectral features for a minimum resolution of $\lambda/\Delta\lambda = 100,000$ are shown in Fig. 3 for $O_2$ and $O_3$.

The high resolution transmission spectra database ($\lambda/\Delta\lambda > 100,000$) is available online www.carlsaganinstitute.org/data and can be used as a tool to optimize our observation strategy, train retrieval methods, as well as interpret upcoming observations with JWST as well as ground-based Extremely Large Telescopes and future mission concepts like Origins, HabEx, and LUOVIR.

**ACKNOWLEDGEMENTS**

The authors acknowledge funding from the Brinson Foundation. and the Carl Sagan Institute